\documentclass[%
 reprint,
 amsmath,amssymb,
 aps,
]{revtex4-1}

\usepackage{graphicx}
\usepackage{dcolumn}
\usepackage{bm}

\usepackage[dvips]{color}

\usepackage[none]{hyphenat}
\usepackage{amsmath}
\usepackage[update,prepend]{epstopdf}
\usepackage{url}
\usepackage[colorlinks=true,
            linkcolor=red,
            citecolor=blue,
            anchorcolor = blue,
            urlcolor=blue]{hyperref}
\usepackage[normalem]{ulem}

\expandafter\def\expandafter\UrlBreaks\expandafter{\UrlBreaks
  \do\a\do\b\do\c\do\d\do\e\do\f\do\g\do\h\do\i\do\j%
  \do\k\do\l\do\m\do\n\do\o\do\p\do\q\do\r\do\s\do\t%
  \do\u\do\v\do\w\do\x\do\y\do\z\do\A\do\B\do\C\do\D%
  \do\E\do\F\do\G\do\H\do\I\do\J\do\K\do\L\do\M\do\N%
  \do\O\do\P\do\Q\do\R\do\S\do\T\do\U\do\V\do\W\do\X%
  \do\Y\do\Z}

\newcommand{\zn}{\mathcal{Z}}

\newcommand{\Be}{{\rm{Be}}}
\newcommand{\Tr}{{\rm{T}}}
\newcommand{\De}{{\rm{D}}}
\newcommand{\beq}{\begin{equation}}
\newcommand{\eeq}{\end{equation}}

\begin{document}
\setlength{\pdfpagewidth}{8.5in}
\setlength{\pdfpageheight}{11in}

\preprint{APS/123-QED}

\title{
A new ICRF scenario for bulk ion heating in D-T plasmas: \\
How to utilize intrinsic impurities in fusion devices in our favour}
\author{Ye.O. Kazakov$^1$, J. Ongena$^1$, D. Van Eester$^1$,
R. Bilato$^2$, R. Dumont$^3$, \\
E. Lerche$^1$, M. Mantsinen$^{4,5}$ and  A. Messiaen$^1$}
\affiliation{%
$^1$ Laboratory for Plasma Physics, LPP-ERM/KMS, EUROfusion Consortium member, Brussels, Belgium \\
$^2$ Max-Planck-Institut f\"ur Plasmaphysik, Garching, Germany \\
$^3$ CEA, IRFM, F-13108 Saint-Paul-lez-Durance, France \\
$^4$ Catalan Institution for Research and Advanced Studies, Barcelona, Spain \\
$^5$ Barcelona Supercomputing Center (BSC), Barcelona, Spain \\
}%


\begin{abstract}
\vspace{-7mm}
A fusion reactor requires plasma pre-heating before the rate of
deuterium-tritium fusion reactions becomes significant.
In ITER, radiofrequency (RF) heating of $^3$He ions, additionally puffed
into the plasma, is one of the main options considered for increasing bulk ion temperature
during the ramp-up phase of the pulse.
In this paper, we propose an alternative scenario for bulk ion heating with RF waves,
which requires no extra $^3$He puff and profits from the presence of intrinsic Beryllium impurities
in the plasma. The discussed method to heat Be impurities in D-T plasmas is shown to provide an even larger fraction of fuel ion heating.

\end{abstract}

\maketitle


\vspace{-5mm}
\noindent
\section{Introduction}
\vspace{-3mm}

In future fusion devices, radio frequency
plasma heating with waves in the ion cyclotron range
of frequencies (ICRF) is the only method capable of
providing a significant fraction of bulk ion heating.
This is due to the fact that electron cyclotron
resonance heating,
by definition, deposits
wave energy to electrons. The same holds for the neutral
beam injection systems because
they will need to operate at rather high beam energies $(\sim 1~{\rm MeV})$
in order to reach the high-density plasma core.

As discussed in \cite{bergeaud2000, eriksson2001, dve2002},
preferential bulk ion heating during the ramp-up phase
has a number of advantages. These include
an improved control over the path to the burn phase
and keeping away from the region where the interaction between
thermal ions and thermal electrons is weak.
As a result, reaching the operational point with higher $Q$ can be done
faster with less auxiliary heating power
than for an electron heating scenario~\cite{bergeaud2000}.
In addition, Tore Supra experiments demonstrated
that higher levels of ion heating were accompanied by
an improved energy confinement~\cite{eriksson2001}.

Applying ICRF power -- somewhat counter-intuitively to its name --  does not necessarily result in
the dominant heating of bulk ions.
In fact, depending on the chosen operational conditions (e.g., ICRF frequency,
antenna phasing, plasma composition), the incoming RF power can be
directly absorbed by various ion species, as well as by electrons. It is the latter condition one
needs to fulfill in order to drive non-inductive current with ICRF system~\cite{iter1999, porkolab1998}.
However, channeling the RF power to be mostly absorbed by ions is not sufficient for
bulk ion heating. In addition, one needs to avoid the formation of a high-energy ion tail due to ICRF.
The critical energy $E_{\rm crit}$ of fast ions, at which
the rate of loss of energy to the electrons and to the ions is equal,
is given by
\mbox{$E_{\rm crit} = 14.8A_{\rm fast}T_{e}\left(\sum_{i} X_{i} Z_{i}^2/A_{i} \right)^{2/3}$}~\cite{stix1972}.
Here, $A_{\rm fast}$ is the atomic mass of the energetic ions, $T_{e}$ is the electron temperature,
$X_{i}=n_{i}/n_{e}$, $Z_{i}$ and $A_{i}$ are the concentration, the charge state and the atomic mass
of the thermal ion species, respectively.
In order to achieve the resultant dominant heating of bulk ions,
the energies of RF-accelerated minority ions should stay in the range of $E_{\rm crit}$
or below.

ICRF scenarios relevant for the D-T plasmas were discussed
in detail in Refs.~\cite{iter1999, bergeaud2000, dve2002, eriksson1999}, with the emphasis
to ITER in, e.g., \cite{budny2012, dumont2013}. Many of these scenarios were studied experimentally during the past
D-T experiments in JET and TFTR tokamaks~\cite{start1998, start1999, rimini2000, wilson1995, phillips1999}.
Currently, the second
harmonic heating of tritium ions $(\omega = 2 \omega_{c \Tr})$ is considered
as the main ICRF scenario for the ITER burn-phase plasmas, and the fundamental
heating of a~small fraction of $^3$He ions $(\omega = \omega_{c, ^{3}{\rm He}})$, additionally puffed into the plasma,
for the ramp-up low-temperature heating stage of the pulse.
In JET, a fourfold increase of the neutron emission was observed when \mbox{$X[^{3}{\rm He}]=4\%$} was injected \cite{start1999}.
The higher reactivity observed with $^3$He minority heating was due to the higher $T_{i}$ reached.
However, currently the supply of $^3$He reduces and the industrial demand of this gas
is progressively increasing~\cite{report2010}.

In this paper, we introduce and discuss a promising alternative bulk ion heating scenario
for the ramp-up phase in D-T plasmas. We will mainly focus on the
applicability of the proposed scheme for ITER. Instead of using additionally puffed
$^3$He ions, we suggest to tune the RF system to heat predominantly intrinsic
Beryllium (Be) impurities, which will be naturally present in ITER plasmas.
Because of the larger atomic mass of Be, such ICRF heating is shown to
provide an even larger fraction of bulk ion heating.
Finally, we discuss how the current design of the ICRF system in ITER can be adapted to exploit
the advantages of beryllium ICRF heating to their maximum.
The proposed method is not restricted to the use of $^{9}{\rm Be}$ only, but other
impurities with a similar charge-to-mass ratio, such as $^{7}{\rm Li}$, $^{22}{\rm Ne}$, $^{40}{\rm Ar}$, etc.
can be used as well. This is discussed in the last section of the paper and is illustrated with a few examples.

\vspace{-3mm}
\section{Three-ion (B{\lowercase{e}})-D-T ICRF scenario}

\vspace{-2mm}
ICRF heating relies on launching the fast magnetosonic wave (FW)
into the plasma by external antennas, typically located
at the low field side (LFS) edge of the plasma. The FW propagation
across the plasma is fairly well described by the well-known dispersion relation~\cite{porkolab1994}
\vspace{-2mm}
\beq
\vspace{-2mm}
n_{\perp, {\rm FW}}^2 \simeq \frac{(\epsilon_{\rm L} - n_{\|}^2)(\epsilon_{\rm R} - n_{\|}^2)}{\epsilon_{\rm S} - n_{\|}^2}
\label{eq:1},
\eeq
where $n_{\perp, \|} = ck_{\perp, \|}/\omega$ is the perpendicular/parallel FW refractive
index, and the tensor components $\epsilon_{\rm S}$, $\epsilon_{\rm L}$ and $\epsilon_{\rm R}$
are those given by Stix~\cite{stix1992}.

The FW power is absorbed by ions, when the wave
crosses the ion cyclotron (IC) resonance $(\omega = \omega_{ci})$ and IC harmonic $(\omega = N \omega_{ci}, N \geq 2)$ layers.
Electron absorption is also possible via a combination
of electron Landau damping
and transit time magnetic pumping,
especially at high electron beta. Electron heating can also occur via mode conversion
of the incoming FW to shorter wavelength modes, which takes place at the
ion-ion hybrid resonance(s) in multi-ion plasmas~\cite{mantsinen2004}.

In general, the FW is elliptically polarized and its electric field can be decomposed as a sum of two components:
the left-hand polarized component $E_{+}$, which rotates in the sense of ions in the magnetic field,
and the right-hand polarized component $E_{-}$ that is aligned with electron rotation.
In his seminal paper~\cite{stix1975}, Stix showed that the absorbed RF power by thermal ions
is almost merely due to the left-hand $E_{+}$ component. One should note here
that it is the plasma, rather than the ICRF system, which defines the wave polarization and
imposes the ratio between $E_{+}$ and $E_{-}$~\cite{stix1975}
\vspace{-2mm}
\beq
\vspace{-2mm}
\left| \frac{E_{+}}{E_{-}} \right| \simeq \left| \frac{\epsilon_{\rm R} - n_{\|}^2}{\epsilon_{\rm L} - n_{\|}^2} \right|.
\label{eq:2}
\eeq
Though the characteristic $n_{\|}$, which appears in Eqs.~(\ref{eq:1}) and (\ref{eq:2}),
can be varied by changing the ICRF antenna phasing and operational frequency, the main contribution to
Eq.~(\ref{eq:2}) generally comes from the tensor elements.

One option for ion absorption in fusion plasmas is applying IC harmonic heating $(N \geq 2)$,
$|E_{+}/E_{-}| \simeq (N-1)/(N+1)$.
This damping mechanism is, however, a finite Larmor radius effect
and hence not appropriate for the beginning of
the heating phase in a reactor (low densities and low temperatures).
Another option -- known as minority heating in two-ion plasmas~\cite{stix1975} -- is to heat a small fraction
of resonant minority ions at their fundamental IC resonance.
This scenario typically shows the best performance at minority concentrations of
$X_{\rm mino} = n_{\rm mino}/n_{e} \approx 3-10\%$ and has been~routinely used for plasma heating.
Though minority heating has a quite strong
single-pass absorption (SPA), especially for hydrogen minority heating
in deuterium majority plasmas,
the ratio of the left- to the right-hand polarization
at the IC resonance of minority ions is limited~\cite{start1999}
\vspace{3mm}
\beq
\vspace{-1mm}
\left| \frac{E_{+}}{E_{-}} \right|_{\omega=\omega_{c2}} \approx \left| \frac{\zn_2 - \zn_1}{\zn_2 + \zn_1} \right| < 1,
\label{eq:3}
\eeq
and the FW is mostly right-hand polarized. Here, we introduced the notation $\zn_{i}=(Z/A)_{i}$, which stands for the ratio of the charge state
to the atomic mass for the ion species. By way of example, for (H)-D heating $|E_{+}/E_{-}| \approx 1/3$,
which is enough to obtain an efficient single-pass absorption at $X_{\rm H} \approx 5\%$
(typical H concentration used in the experiments). Yet, two-ion ICRF
minority heating at much lower minority concentrations (e.g., $X_{\rm mino} \ll 1\%$)
is usually characterized by very poor
wave absorption and for this reason is not often used in practice.

In our recent paper \cite{kazakov2015}, we presented a method
to maximize the $E_{+}$ component and, subsequently,
to achieve a strong ion absorption at very low minority concentrations ($\sim 1\%$ or lower). This
enhanced ion absorption requires the presence of three ion species in a~plasma
(with a different $Z/A$): two main ion species (X~and Y) and a third resonant ion species (Z), present at a very small concentration.
The key idea is to locate the left-hand polarized fast wave L-cutoff, which is defined by the condition $\epsilon_{\rm L}=n_{\|}^2$
and which largely enhances the associated RF field component $E_{+}$,
close to the fundamental cyclotron resonance of the minority ion species Z.
This matching can be achieved by adjusting the density ratio between the two main ion species Y:X.

Note that not all ICRF scenarios with three ion species can lead to $E_{+}$ enhancement.
One of the necessary conditions is that
\vspace{-1mm}
\beq
\vspace{-1mm}
\min\{\zn_1, \zn_2\} < \zn_{3} < \max\{\zn_1, \zn_2\},
\label{eq:5}
\eeq
i.e. the resonant minority ions, one aims to heat, should have a cyclotron
resonance located between the IC resonances of the two bulk ion species.
The main ICRF scenario for bulk ion heating in D-T ITER plasmas, viz. ($^{3}$He)-DT,
does not belong to this list since $\zn_3 > \zn_{1,2}$.

\begin{figure*}
\includegraphics[trim=0cm 0cm -0mm -0.0cm, clip=true, height=0.36\textwidth]{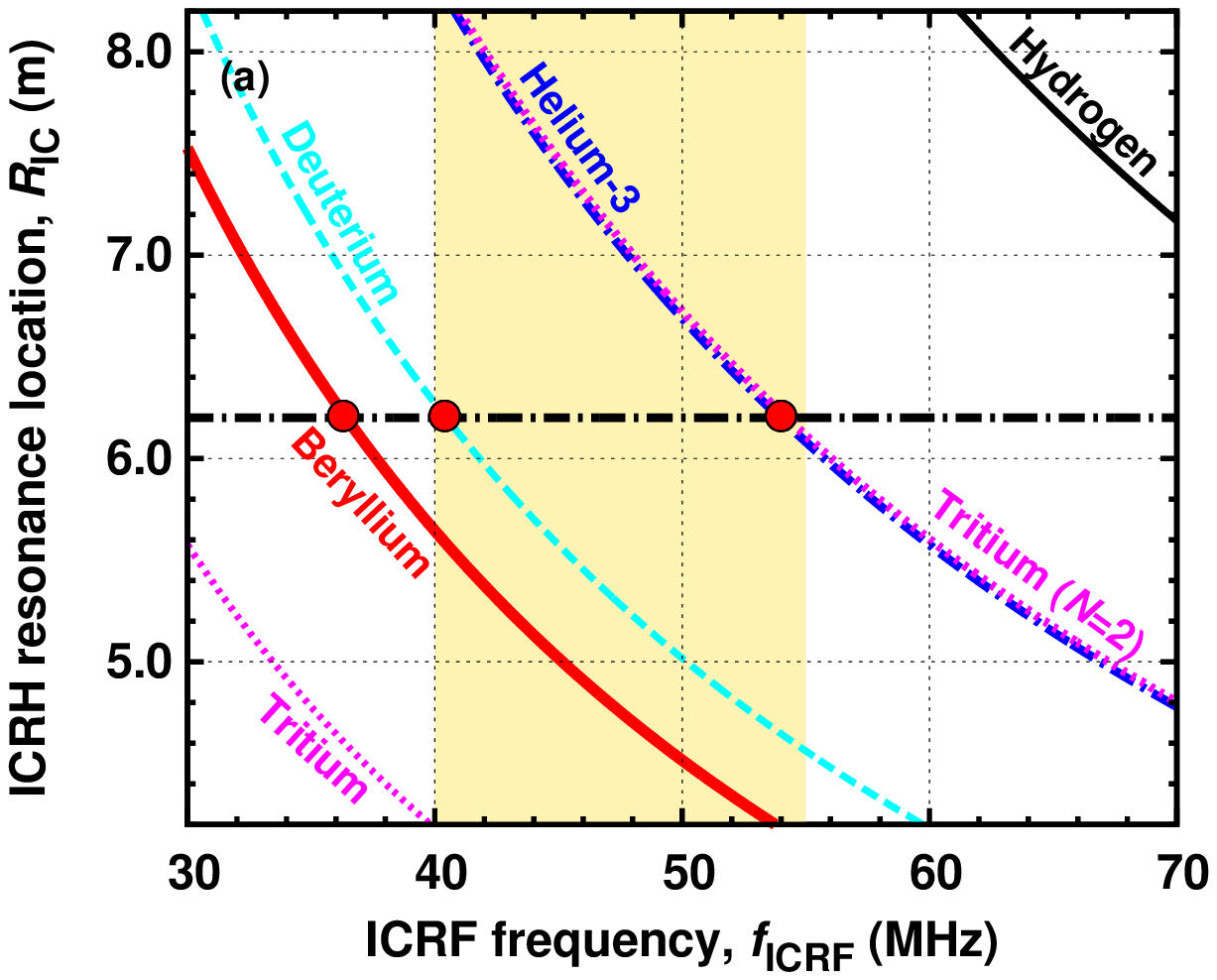}
\includegraphics[trim=0cm 0cm -0mm -0.0cm, clip=true, height=0.36\textwidth]{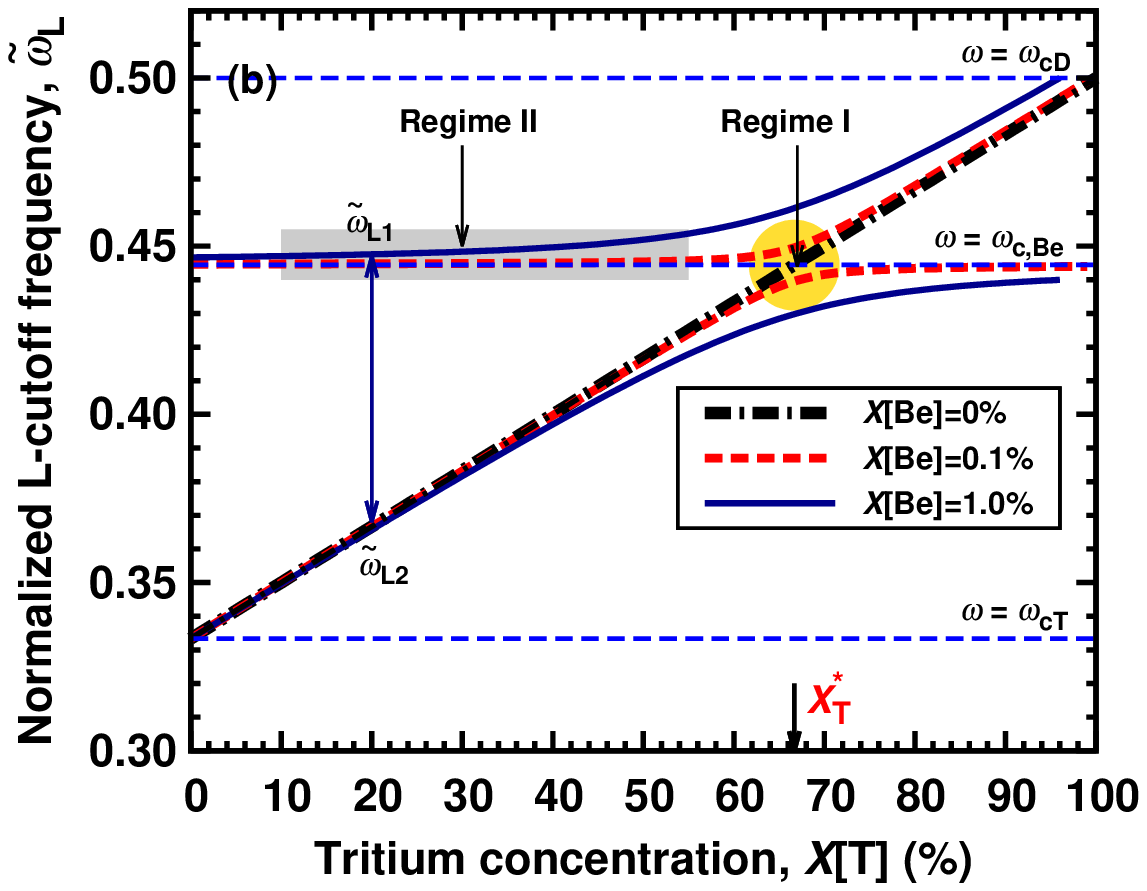}
\vspace{-5mm}
\caption{
(a) Location of the IC resonances in the ITER plasma at the full magnetic field ($B_0 = 5.3~{\rm T}$).
The official ITER frequency range for the ICRF system $f=40-55~{\rm MHz}$ is highlighted.
(b) The normalized L-cutoff frequency, $\tilde{\omega}_{\rm L} = \omega_{\rm L}/\omega_{\rm cH}$ as
a function of tritium and beryllium concentrations in a (Be)-D-T plasma.}
\label{fig1}
\end{figure*}

Tungsten and beryllium have been chosen as the plasma-facing components
for ITER. Accordingly, the carbon wall was replaced with the ITER-like wall in JET
since 2009. Therefore, JET and ITER plasmas will
unavoidably include a non-negligible amount
of Be impurities. For example, $X_{\rm Be}=1-2\%$ is estimated as the
background Be level in JET plasmas~\cite{brezinsek1, brezinsek2}:
the measured core $Z_{\rm eff} \approx 1.2$ yields $X_{\rm Be} \simeq 1.4\%$~for
the residual carbon content of about $X_{\rm C} \simeq 0.1\%$
(note that $Z_{\rm eff} \approx 1.1$ yields $X_{\rm Be} \approx 0.6\%$).
A~charge-to-mass ratio of the fully-ionized beryllium ions ($A=9$, $Z=4$) satisfies
the condition of Eq.~(\ref{eq:5}),
$\zn_{\rm T} < \zn_{\rm Be} < \zn_{\rm D}$, and therefore (Be)-D-T scenario
belongs to the set of three-ion ICRF heating scenarios. Though
in Ref.~\cite{wilson1998} it was correctly outlined that the application of
D-T mode conversion heating could be hampered due to the Be presence,
our recent results~\cite{kazakov2015} show that
the impact of the three-ion ICRF scenarios can be reverted and
such scenarios have a great potential for fusion research
(fast-ion generation and bulk plasma heating).
Below, we explore the features
of the (Be)-D-T ICRF scenario in ITER-like plasmas and compare
the potential of bulk ion heating for this scenario with
the more traditional ($^3$He)-DT scheme.

\vspace{-3mm}
\section{(B{\lowercase{e}})-D-T heating in ITER}
\vspace{-2mm}
Figure~\ref{fig1}(a) shows the location of the main IC resonances in ITER ($R_0$\,=\,6.2~m, $a$\,=\,2.0 m)
as a~function of the ICRF frequency for the
nominal magnetic field $B_0=5.3~{\rm T}$.
In ITER, two ICRF antennas mounted in mid-plane ports are envisaged
to operate within the range $f=40-55~{\rm MHz}$~\cite{lamalle2009, wilson2015}.
Such a frequency range for ITER provides central $N=1$
and \mbox{$N=2$} heating for $^3$He and T ions at $f \approx 53~{\rm MHz}$,
and fundamental IC heating of D ions at $f \approx 40~{\rm MHz}$
(for (D)-T pulses). We also depict the location
of the IC resonance of Be impurities (red solid line) in Fig.~\ref{fig1}. One can see that
the distance between the Be and D cyclotron resonances is about $70~{\rm cm}$,
and a frequency only slightly below the official minimum ICRF frequency for ITER
$f \approx 36-37~{\rm MHz}$ is required to locate the Be resonance close to the plasma center.
A frequency $f=37~{\rm MHz}$ is used throughout~the paper, unless otherwise stated.
According to the arguments presented in the last section of this paper,
the ICRF system in ITER can operate at this frequency,
without any dramatic changes to the present design.
As will also be discussed below, operation at $f=40~{\rm MHz}$ can still
provide dominant Be heating at the price of yielding reasonably off-axis power deposition.

\begin{figure*}
\includegraphics[trim=0cm 0cm -0mm -0.0cm, clip=true, width=0.36\textwidth]{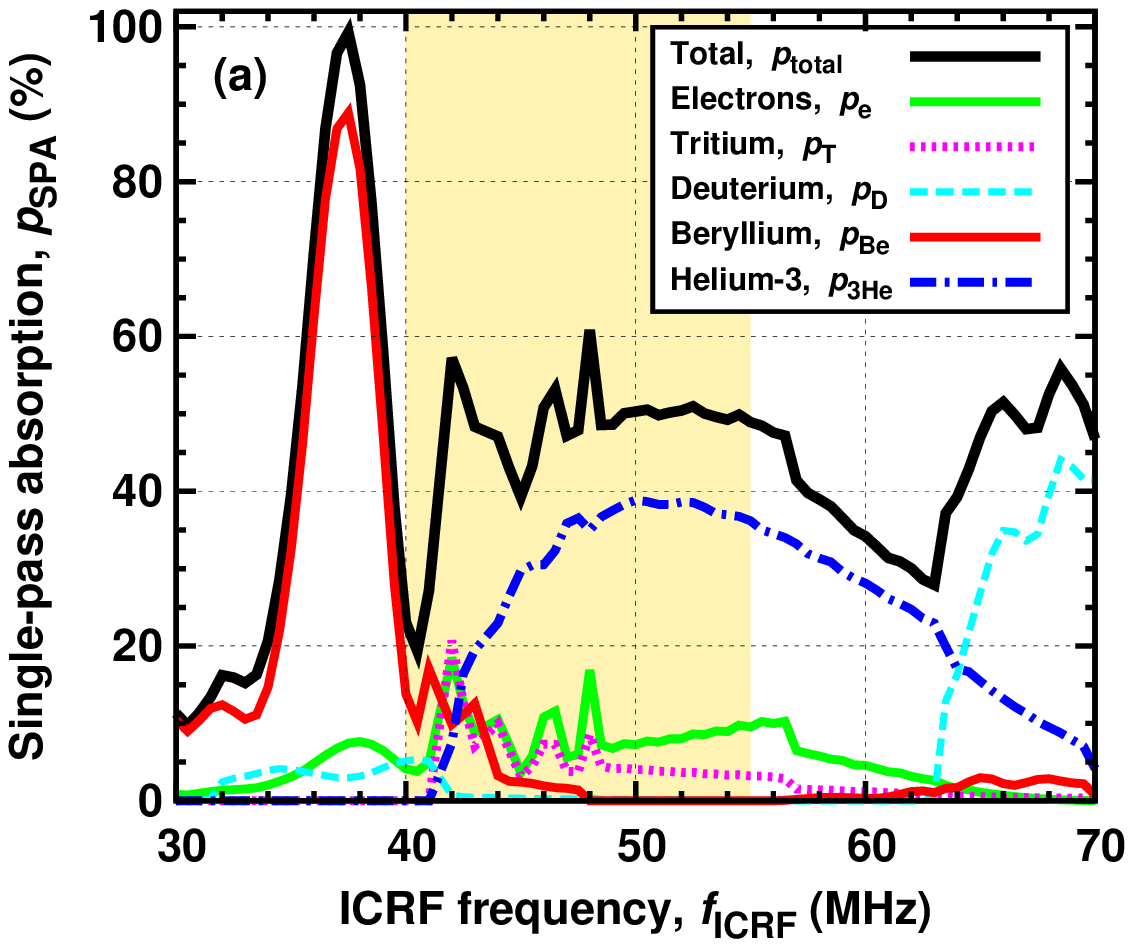}
\includegraphics[trim=0cm 0cm -0mm -0.0cm, clip=true, width=0.36\textwidth]{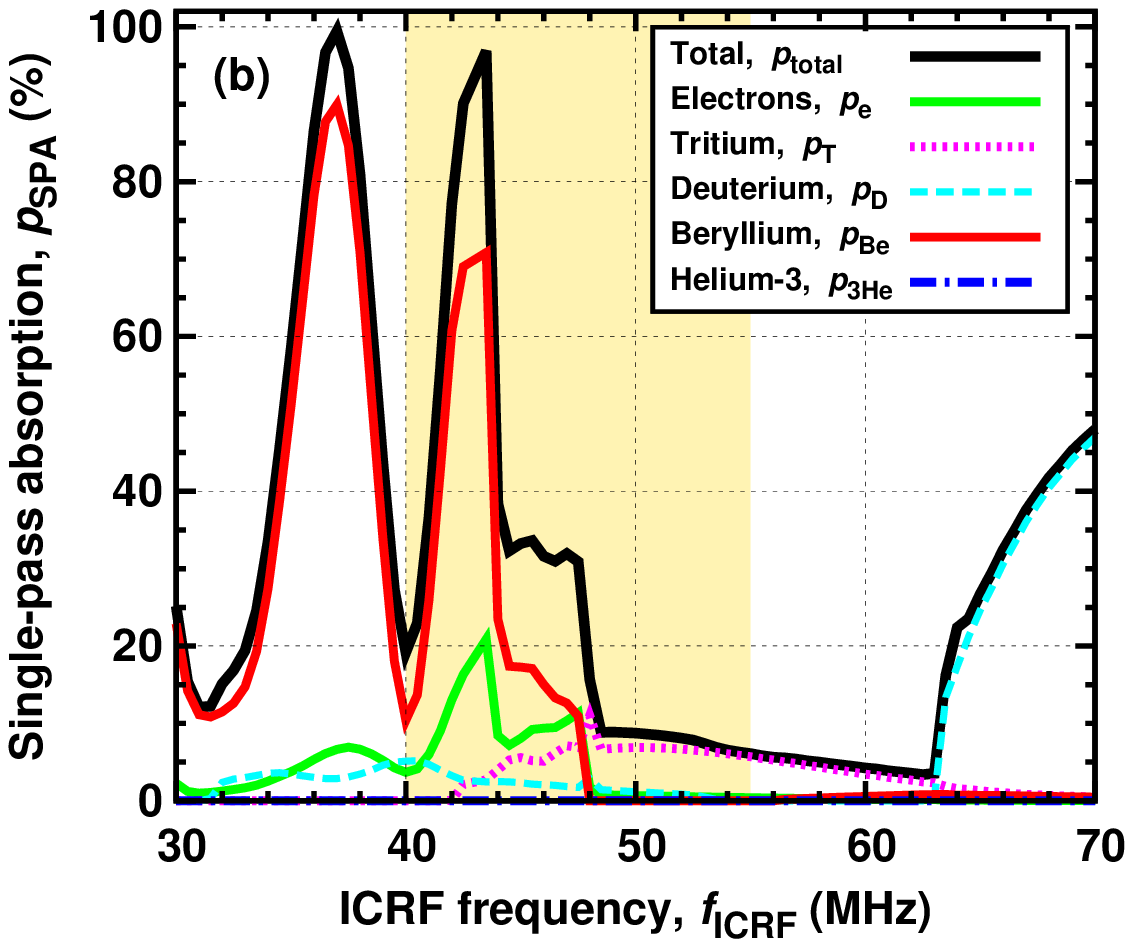}
\vspace{-5mm}
\caption{Single-pass absorption coefficients as a function of the ICRF frequency
for the ITER ramp-up phase as computed by TOMCAT ($B_0 = 5.3~{\rm T}$, D:T=1:1, $T_{0}=4~{\rm keV}$,
$n_{e0} = 6\times 10^{19}~{\rm m}^{-3}$, $k_{\|}^{\rm (ant)}= 3~{\rm m}^{-1}$):
(a) ($^{3}$He)-DT scenario ($X_{^{3}{\rm He}}=2\%$, $X_{\rm Be}=1\%$);
(b)~(Be)-D-T scenario (no $X_{^{3}{\rm He}}$, $X_{\rm Be}=1\%$).}
\label{fig2}
\end{figure*}

Depending on the impurity concentration, one can distinguish two different regimes
of the three-ion ICRF heating.
If the concentration of Be ions
is extremely small $(X_{\rm Be} \ll 1\%)$,
the optimal T concentration for maximizing Be absorption
can be found from Eq.~(7) of Ref. \cite{kazakov2015}, viz.
$X_{\rm T}^{*} \approx (\zn_{\rm Be} - \zn_{\rm T})/(\zn_{\rm D} - \zn_{\rm T}) \approx 67\%$
(D:T $\simeq$ 1:2). At such $X_{\rm T}$, the L-cutoff in D:T plasmas (see dashed-dotted line in Fig.~\ref{fig1}(b))
and the cyclotron resonance of Be impurities intersect.
However, in three-ion Be-D-T plasmas with a finite $X_{\rm Be}$,
the L-cutoff can not cross the IC resonance of Be.
This is clear from the equation for the normalized L-cutoff frequency
\mbox{$(\tilde{\omega}_{\rm L}=\omega_{\rm L}/\omega_{c{\rm H}})$},
which can be written as follows~\cite{kazakov2013} \\
\vspace{-1mm}
\beq
\frac{X_{\rm T}}{\tilde{\omega}_{\rm L} - \zn_{\rm T}} + \frac{X_{\rm D}}{\tilde{\omega}_{\rm L} - \zn_{\rm D}} + \frac{4X_{\rm Be}}{\tilde{\omega}_{\rm L} - \zn_{\rm Be}} \simeq 0.
\label{eq:6}
\eeq
Note that any solution of Eq.~(\ref{eq:6}) can be translated into the corresponding radial coordinate of the L-cutoff
via \mbox{$R_{\rm L} \approx R_0\,\tilde{\omega}_{\rm L}\,15.25\,B_0({\rm T})/f({\rm MHz})$}.
In fact, as follows from Fig.~\ref{fig1}(b), in the vicinity of $\omega=\omega_{c, {\rm Be}}$, the curve for $\tilde{\omega}_{\rm L}$
splits in two, and generally there are two meaningful solutions of Eq.~(\ref{eq:6}) at a given $X_{\rm T}$ and $X_{\rm Be}$.
Far from the region of the resonant crossing ($X_{\Tr} \approx X_{\Tr}^{*}$), they can be approximated with
expressions
\beq
\vspace{-1mm}
\begin{split}
& \tilde{\omega}_{\rm L1} \approx \zn_{\Be} + \frac{4(\zn_{\De} - \zn_{\Be})X_{\Tr}^{*}X_{\Be}}{X_{\Tr}^{*} - X_{\Tr}}, \\
& \tilde{\omega}_{\rm L2} \approx \zn_{\Tr} + (\zn_{\De} - \zn_{\Tr})X_{\Tr} - \frac{4(\zn_{\De} - \zn_{\Be})X_{\Tr}X_{\Be}}{X_{\Tr}^{*} - X_{\Tr}}.
\end{split}
\label{eq:7}
\eeq
In the vicinity of $X_{\Tr} \approx X_{\Tr}^{*}$, these approximations
are not valid. For very low $X_{\Be} \ll 1\%$,
the gap between two solutions $\tilde{\omega}_{\rm L1}$ and $\tilde{\omega}_{\rm L2}$ is small,
and the optimal conditions for the ICRF heating
occur at the tritium concentration $X_{\Tr}^{*}$, at which the two asymptotics intersect.
We have depicted such conditions as `Regime~I' in Fig.~\ref{fig1}(b), and this
heating regime has been suggested for fast-ion generation with ICRF
and studied in detail in \cite{kazakov2015}.

However, for higher Be concentrations ($X_{\Be} \sim 1\%$),
at $X_{\Tr} \simeq X_{\Tr}^{*}$ the gap
between $\omega_{\rm L1}$ and $\omega_{c, {\rm Be}}$ is already quite large.
As a result, the distance between the region with the enhanced $E_{+}$
and Be cyclotron resonance exceeds the Doppler width of the IC
resonance, and minority ion absorption at such conditions significantly decreases.
Thus, as already noted in \cite{kazakov2015}, the optimal tritium
concentration for maximizing Be absorption shifts towards
lower values of $X_{\Tr}$. These conditions, which are schematically
depicted as `Regime~II' in Fig.~\ref{fig1}(b), correspond to the conditions
of (Be)-D-T heating in ITER and will be studied in this paper.

We start with a comparison between the ($^3$He)-DT and (Be)-D-T ICRF scenarios
by evaluating the single-pass absorption coefficients
(a fraction of the incoming RF power absorbed by plasma species
as the wave propagates from the
LFS edge to the high field side edge)
as a function of ICRF frequency.
These are computed with the TOMCAT code~\cite{tomcat} and are depicted in Fig.~\ref{fig2}.
For both scenarios, the background \mbox{$X_{\Be}=1\%$} and an optimal D:T=1:1 ratio is assumed.
For the $^3$He minority scenario, shown in Fig.~\ref{fig2}(a), an additional $X_{^{3}{\rm He}}=2\%$ is considered.
Because of the plasma dilution, the concentration of fuel D-T ions for this scenario
is somewhat lower than for Be minority heating \mbox{($X_{\De, \Tr}=46\%$ vs. $X_{\De, \Tr}=48\%$)}.
We focus on the ITER ramp-up phase, when the plasma density and temperatures
are rather low and the rate of D-T fusion reactions is small, such that
the contribution of alpha-particles to the power balance is negligible.
As~baseline conditions, we consider $T_0 = 4~{\rm keV}$, \mbox{$n_{e0}=6\times 10^{19}~{\rm m}^{-3}$},
$n_{\rm tor}=27$ ($k_{\|}^{\rm (ant)}=3~{\rm m}^{-1}$). Such a~value of the FW parallel wavenumber is representative
for the $[0;0;\pi;\pi]$ toroidal phasing of the ITER ICRF antenna~\cite{dumont2009, budny2012}.
An influence of the antenna toroidal spectrum on ICRF heating scenarios in ITER was
studied in \cite{dumont2009, dumont2013}. For the ($^3$He)-DT scenario, two modelling approaches
(full-spectrum computations and retaining only the dominant toroidal mode number)
were shown to give almost identical results (see Table II of Ref.~\cite{dumont2009}).
Hence, for proof-of-principle computations which we present in this paper,
we have adopted a simpler approach and considered only the dominant wavenumber.

\begin{figure*}
\includegraphics[trim=0cm 0cm -0mm -0.0cm, clip=true, width=0.30\textwidth]{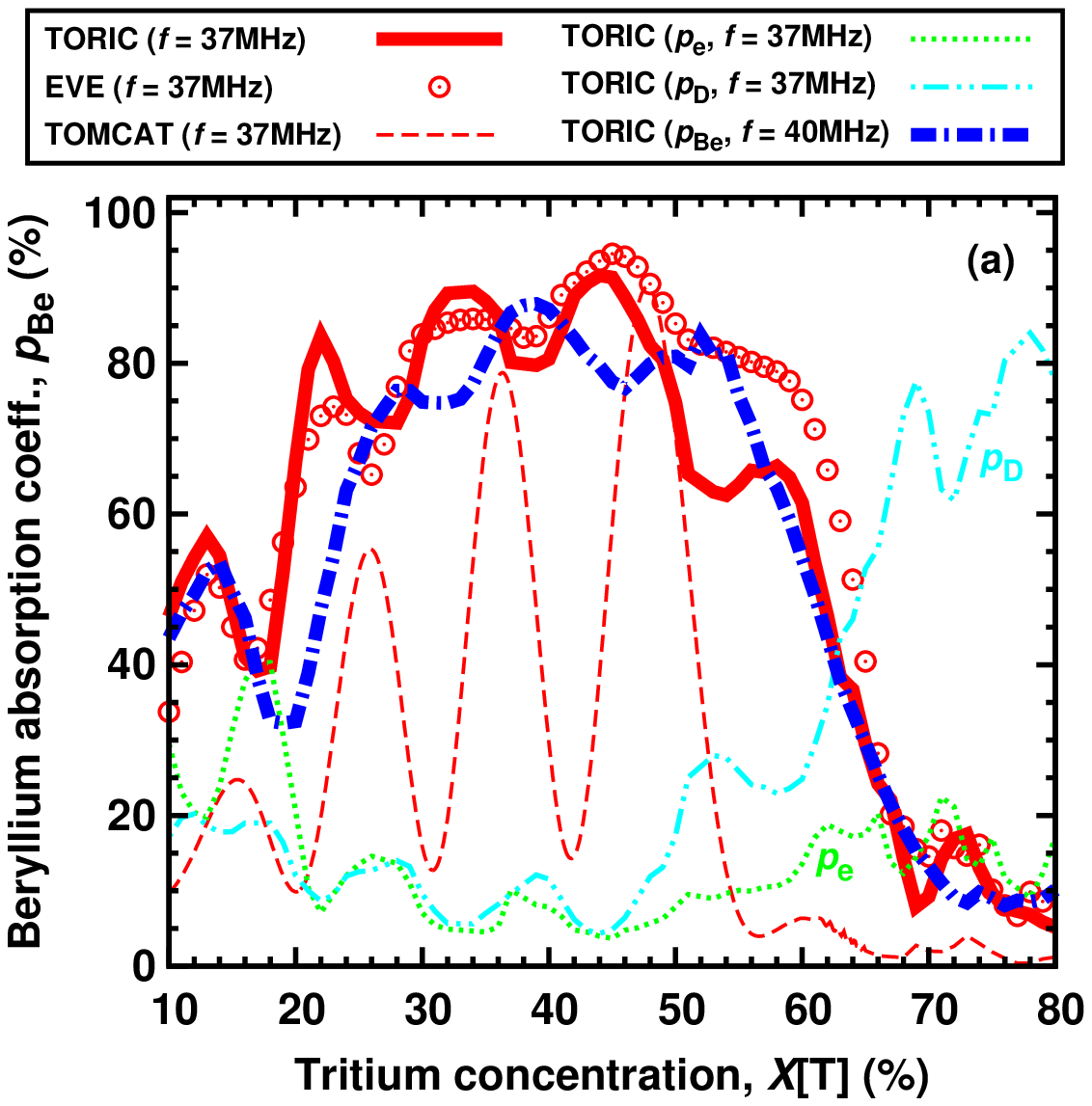}
\includegraphics[trim=0cm 0cm -0mm -0.0cm, clip=true, width=0.34\textwidth]{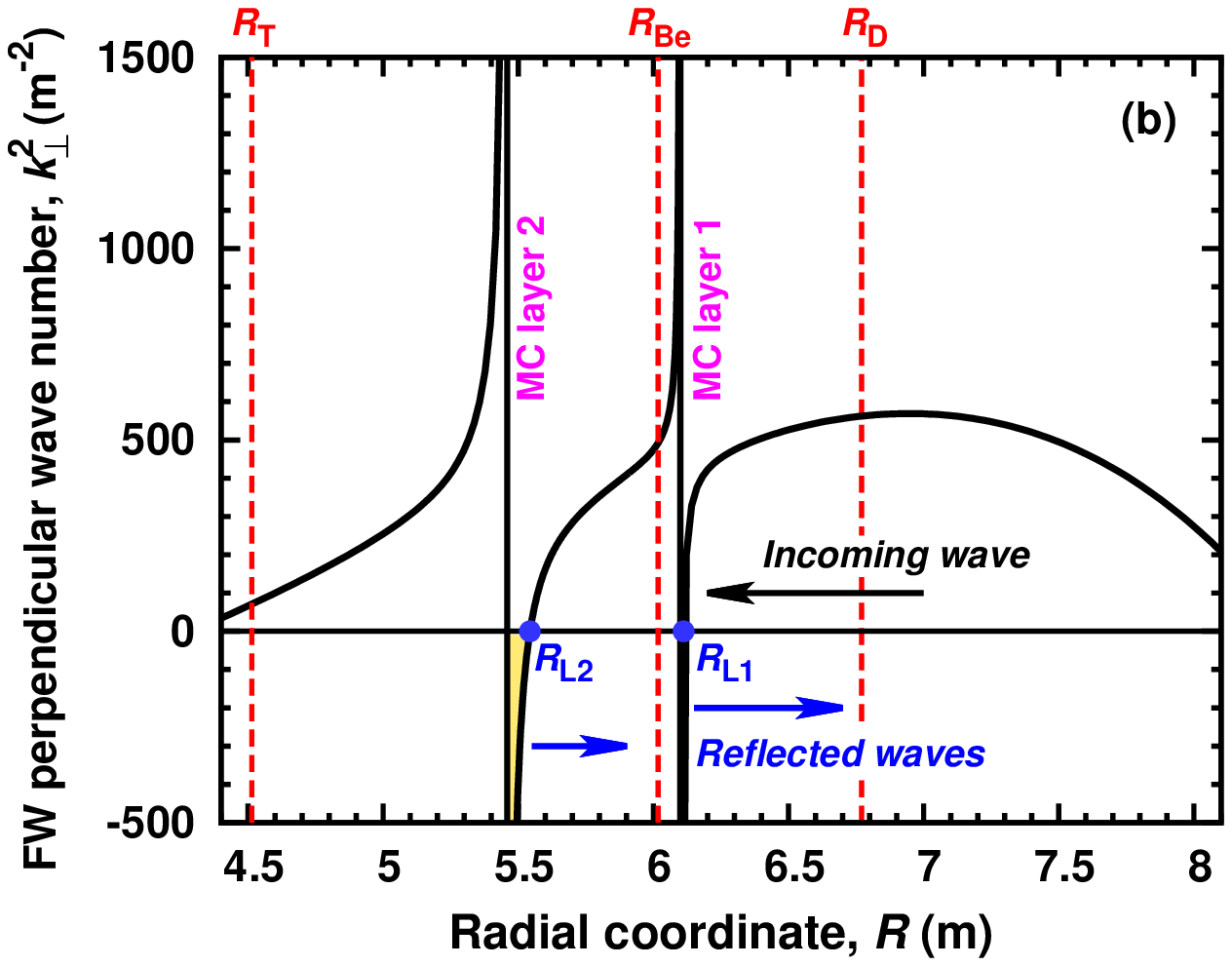}
\includegraphics[trim=0cm -0.0cm -0mm -0.0cm, clip=true, width=0.30\textwidth]{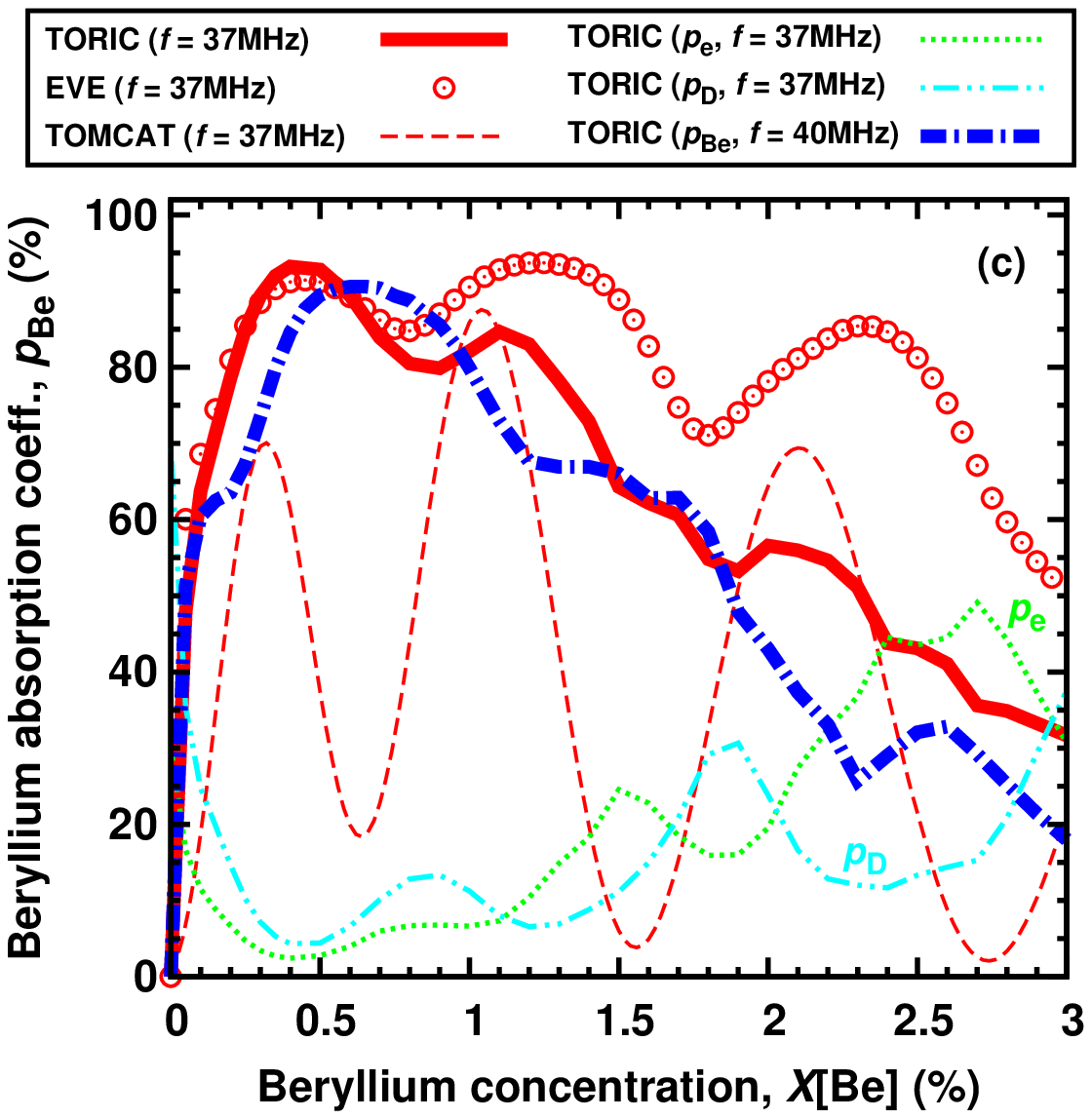}
\vspace{-1mm}
\caption{Beryllium impurity ICRF heating scenario, (Be)-D-T: (a) Absorption coefficients as
a function of tritium concentration ($X_{\Be}=1\%$), as~computed by TORIC, EVE and TOMCAT;
(b) Radial dependence of the real part of $k_{\perp, {\rm FW}}^2$ ($X_{\Be}=1\%$, $X_{\De}=X_{\Tr}=48\%$);
(c) $p_{\rm abs}$ vs. beryllium concentration (D:T=1:1).}
\label{fig3}
\end{figure*}

As follows from Fig.~\ref{fig2}, for the considered conditions the single-pass absorption for the $^3$He minority scenario
is $p_{\rm ^{3}He} \approx 40\%$ at $f \approx 50~{\rm MHz}$.
The single-pass absorption for the Be heating scenario can be stronger
\mbox{$p_{\rm Be} \approx 90\%$}, if operating at a lower frequency \mbox{$f \approx 37~{\rm MHz}$}.
Note that $p_{\rm ^{3}He} \approx 40\%$, which is computed for the ($^3$He)-DT scenario,
does not mean poor ICRF heating. Typically, such $p_{\rm SPA}$ is still high enough for having good
multi-pass ICRF performance ($p_{\rm SPA}=40-50\%$ reflects that the FW needs to pass the resonant
layer several times before the power is absorbed).
Note that operation at \mbox{$f \approx 43~{\rm MHz}$}
(secondary peak of Be absorption in Fig.~\ref{fig2}(b))
is not very promising since then the RF power is deposited
appreciably off-axis $((r/a)_{\Be} \approx 0.5$, cf. Fig.~\ref{fig1}(a)).

There is another criterion to consider: one needs to make sure that
the ICRF scheme in question is robust with respect to changes in the D:T ratio and Be
concentration.
Since the results shown in Fig.~\ref{fig2} were computed with a 1D code, which neglects
a number of effects intrinsic for the tokamak geometry such as a finite
poloidal magnetic field, a more sophisticated modelling with the 2D full-wave codes
TORIC~\cite{toric1, toric2} and EVE~\cite{eve}
was done to
increase the degree of realism.
Note that contrary to the 1D computations,
in 2D modelling (since the tokamak vessel acts as a Faraday cage)
all the RF power is eventually absorbed in the plasma,
and the computed absorption coefficients reflect the relative strength of various damping mechanisms.
For the 2D computations we adopted a Shafranov shift of $10~{\rm cm}$ and
applied $B_{0}(R_{0}=6.2~{\rm m}) = 5.4~{\rm T}$
such that the magnetic field at the magnetic axis is $5.3~{\rm T}$ as for the 1D computations.
There is generally a good agreement between TORIC and EVE results.
Figure~\ref{fig3}(a) shows that damping of the RF power on Be impurities is the dominant absorption channel
$(p_{\rm Be} > 60\%)$ for a wide range of tritium
concentrations, namely $X_{\Tr}=20-60\%$. This is in line with the arguments behind Fig.~\ref{fig1}(b)
and discussions above. The remaining incoming RF power is almost equally split
between fuel D ions and electrons.

One can also notice a modest oscillatory behavior of $p_{\rm Be}$ in Fig.~\ref{fig3}(a).
TORIC computations predict that the maxima of Be absorption are reached at $X_{\Tr} \approx 22\%, 34\%$, $44\%$
and $58\%$, and minima at $X_{\Tr} \approx 28\%, 39\%$ and $54\%$, respectively.
This occurs due to a constructive/destructive interference effect.
Similar to the structure considered in \mbox{\cite{mayoral2006, lamalle2006, dve2012, kazakov2010}},
for (Be)-D-T plasmas there are two MC layers located in the plasma (see Fig.~\ref{fig3}(b)).
As noted, Eq.~(\ref{eq:6}) has two meaningful solutions: the first L-cutoff $R_{\rm L1}$ is located
close the IC resonance of Be ions, $R_{\rm Be}$ (to the lower magnetic field side of that)
and the second $R_{\rm L2}$ to the higher field side of $R_{\rm Be}$.
In contrast to `Regime~I' heating, when the impurity concentration
is too small to form the two well-separated MC layers, for this regime
the RF power is not very efficiently absorbed,
when the FW passes through the first MC layer and Be cyclotron resonance $(p_{1} \simeq 30\%)$.
The transmitted FW ($\mathcal{T} \simeq 40\%$) reaches the second MC layer
and is almost entirely reflected back at $R_{\rm L2}$.
As a result, the interference of the two reflected waves determines the resultant
reflection and absorption coefficients. Analytical formulae
describing the constructive/destructive interference effect for minority heating regime
were derived in~\cite{kazakov2012}. According to Eqs.~(5) and (6) of \cite{kazakov2012},
the minority ion absorption varies between $p_{i, \rm min} \approx 13\%$ and $p_{i, \rm max} \approx 99\%$,
depending on the phase difference $\Delta \phi$ between the two reflected waves.
An average single-pass absorption coefficient is then
$\bar{p}_{i}=p_{1} + \mathcal{T}(1-\mathcal{T}) \approx 54\%$.

The two MC layers shown in Fig.~\ref{fig3}(b) have essentially different radial width.
Whereas for the first layer $\Delta_{\rm MC1} \simeq 1-1.5~{\rm cm}$,
the second MC layer is much wider $\Delta_{\rm MC2} \simeq 10~{\rm cm}$ and is, thus, non-transparent
for the wave propagation.
As the concentration of T ions
is gradually decreased from $X_{\Tr}^{*}$, the distance between
$R_{\rm L2}$ and the first MC layer increases (see Eq.~(\ref{eq:7}))
\vspace{-1mm}
\begin{multline}
\nonumber
\Delta R_{12} \simeq \left[(\zn_{\Be} - \zn_{\Tr}) - (\zn_{\De} - \zn_{\Tr})X_{\Tr}\right]R_{\Be}/\zn_{\Be} \approx \\
\approx \left(1 - 3X_{\Tr}/2\right)R_0/4.
\end{multline}
Every new maximum/minimum shown in Fig.~\ref{fig3}(a) marks a case, when
the phase term
$\Phi \simeq \bar{k}_{\perp, \rm FW} \Delta R_{12}$ has changed by $\delta \Phi = \pi/2$,
i.e. the radial distance $R_{12}$
has increased roughly by a~quarter of the average FW perpendicular wavelength.
Naturally,~such oscillations
are much more pronounced in 1D (TOMCAT) computations, and the modulation amplitude
for 2D $p_{\rm abs}$--computations is much smaller.
Since $\bar{k}_{\perp, {\rm FW}} \simeq (\omega_{\rm pH}/c)\zn_{\Be}\left( \sum_{i} X_{i}A_{i} \right)^{1/2} \approx 0.7(\omega_{\rm pH}/c)$~with
\mbox{$\omega_{\rm pH}=(4\pi n_{e} e^2/m_{\rm H})^{1/2}$},
a characteristic tritium variation, marking the transition from the maximum of $p_{\Be}$ to the minimum
and vice versa,
has been derived
\beq
\delta X_{\rm T} \simeq \frac{8\pi}{3\bar{k}_{\perp} R_0} \approx 0.27/(\sqrt{n_{e,20}}\,R_0\left({\rm m})\right).
\label{eq:8}
\eeq
Here, $n_{e, 20}$ is the plasma density expressed in the units
$10^{20}~{\rm m}^{-3}$ and $R_0$ is the major radius of the machine in meters.

\begin{figure*}
\includegraphics[trim=0cm 0cm -0mm -0.0cm, clip=true, width=0.48\textwidth]{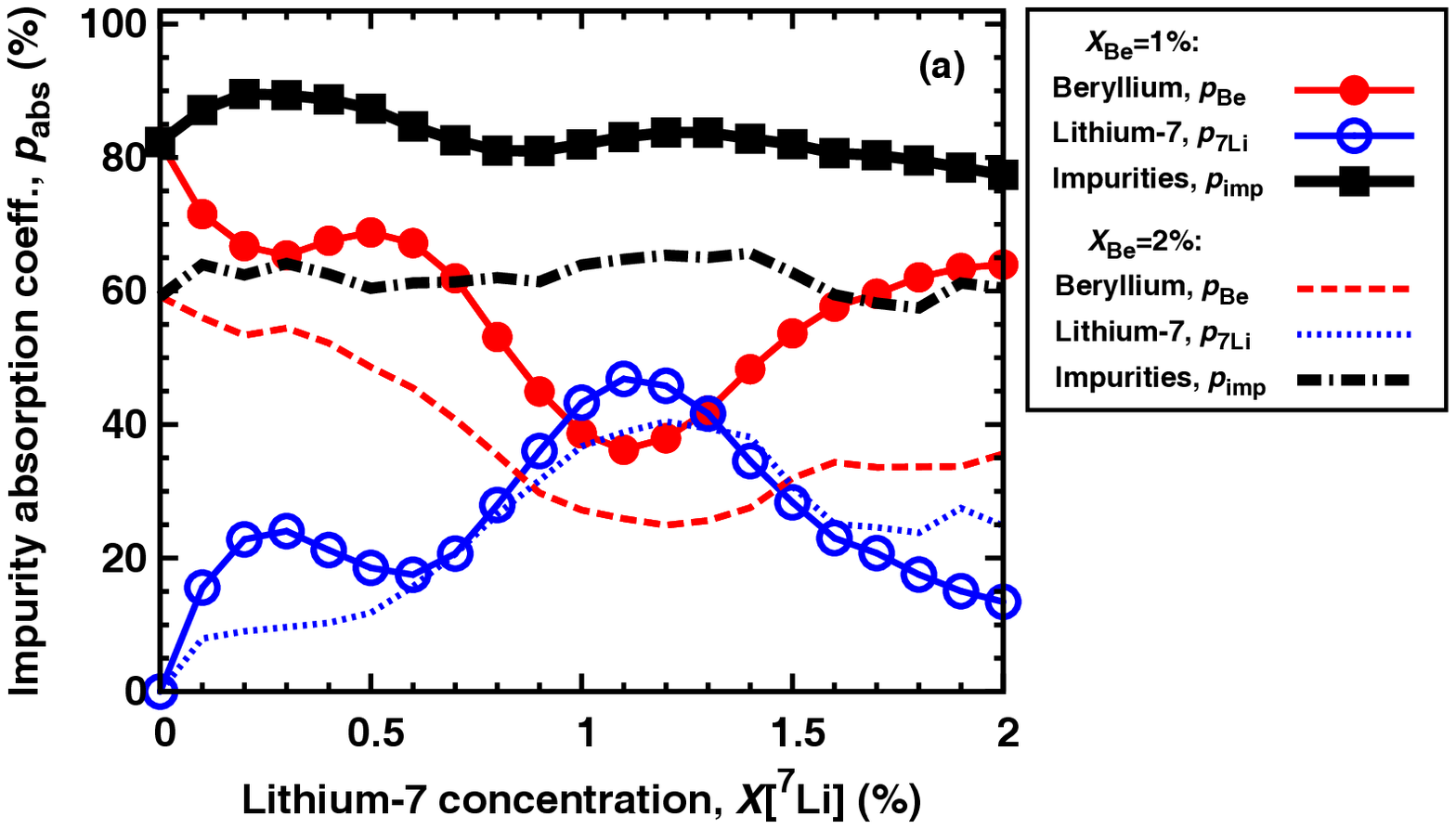}
\includegraphics[trim=0cm 0cm -0mm -0.0cm, clip=true, width=0.48\textwidth]{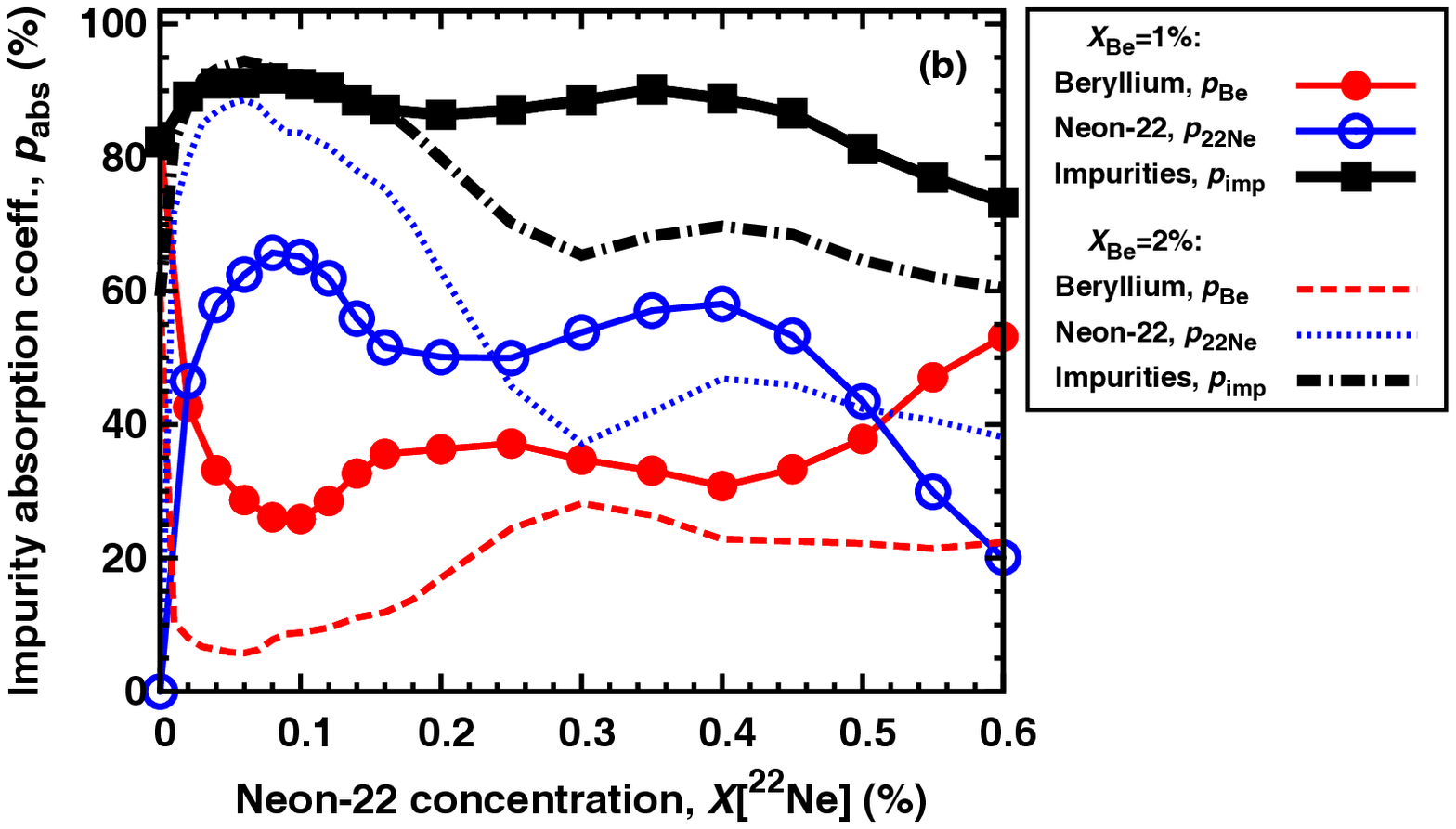}
\vspace{-4mm}
\caption{An effect of extra impurity species with $Z/A \approx 0.43-0.45$: absorption coefficients as a function of
(a) $^{7}$Li and (b)~$^{22}{\rm Ne}$ concentrations. Lines with symbols correspond to the case $X_{\rm Be}=1\%$
(data for $X_{\rm Be}=2\%$ case is shown with lines only).
}
\label{fig4}
\end{figure*}

Accurate measurements of the concentration of Be impurities in the plasma center
are not readily available. Though we fixed $X_{\Be}=1\%$ in our previous
computations, one should consider this value only as a rough estimate.
Figure~\ref{fig3}(c) depicts the sensitivity of the 2D absorption coefficients as
a function of $X_{\Be}$. The concentrations of fuel D and T ions for every
numerical run were
adjusted accordingly to obtain an optimal D:T=1:1 mixture.
Our results show that the RF power is absorbed mostly by Be impurities
within a~fairly wide range $X_{\Be} \approx 0.1\%-2\%$. For the considered conditions, the Be~absorption 
has peaks around \mbox{$X_{\rm Be} \approx 0.5\%$} and $1.2\%$, and tends to decrease at higher concentrations.
At~\mbox{$X_{{\rm Be}} \gtrsim 2.4\%$}, electron
absorption starts to be equally important. Should ITER plasmas include high Be levels $X_{\rm Be} \gtrsim 2.5-3\%$,
this will result in a~reduced Be~absorption.
However, under these conditions at least 20\% of the output fusion power will be lost
due to the fuel dilution.
A~general conclusion, which one can draw from studying Figs.~\ref{fig3}(a) and (c),
is that for the given ICRF frequency \mbox{$f=37~{\rm MHz}$}
Be~absorption is the dominant channel of the RF absorption
for a wide range of realistic plasma conditions in ITER.

The dashed-dotted lines shown in Fig.~\ref{fig3}(a)~and~(c) depict
the absorption coefficient by Be impurities computed with TORIC for $f=40~{\rm MHz}$.
It is clear that Be absorption dominates over the other damping mechanisms
for this ICRF frequency as well.

The presence of high-$Z$ impurities like tungsten has
a~small impact on the (Be)-D-T scenario. A~typical charge state for W
in the plasma core is \mbox{$Z \simeq 40-50$}~\cite{putterich},
and such
impurities have $(Z/A)_{\rm W} \approx 0.2-0.3$,
which is very different to $(Z/A)_{\rm Be}$.
On the contrary, impurities with a~charge-to-mass ratio close to that of Be ions, e.g. $^{7}{\rm Li}^{3+}$,
could have a non-negligible effect on RF power absorption.
Figure~\ref{fig4}(a) shows absorption coefficients by Be and $^{7}{\rm Li}$ impurities computed by TORIC for D:T=1:1 plasmas.
As follows from this figure, the total absorption of RF power by
two different impurity species $p_{\rm imp}=p_{\Be}+p_{^{7}{\rm Li}}$ remains nearly constant.
Since $Z/A$ for Be and $^{7}{\rm Li}$ ions is not identical,
the IC resonant layers of these impurities are separated by a distance about 20~{\rm cm},
and therefore the radial position of the maximum of impurity absorption
is somewhat different (peaking at $r/a \simeq 0.1$ and 0.2, respectively).

\begin{table}
\vspace{-4mm}
\caption{\label{tab1}
Location of IC resonant layers $\omega=\omega_{ci}$
for bulk ion and impurity species, $x_{\rm IC} = R_{\rm IC} - R_0$ ($R_0=6.2~{\rm m}$, \mbox{$a=2.0~{\rm m}$}).}
\vspace{2mm}
\begin{tabular}{|l|c|c|c|c|c|c|}
\hline
 & \,Ion species & T & $^7$Li & Be & $^{22}$Ne & D, $^{20}$Ne \\
\hline
$37~{\rm MHz}$  & $x_{\rm IC}\,({\rm m})$ & \,-1.69\, & \,-0.40\, & \,-0.19\, & \,-0.05\, & \,0.57\, \\
& $x_{\rm IC}/a$ & \,-0.85\, & \,-0.20\, & \,-0.09\, & \,-0.03\, & \,0.28\, \\
\hline
$40~{\rm MHz}$ & $x_{\rm IC}\,({\rm m})$ & \,-2.03\, & \,-0.84\, & \,-0.64\, & \,-0.51\, & \,0.06\, \\
& $x_{\rm IC}/a$ & \,-1.01\, & \,-0.42\, & \,-0.32\, & \,-0.26\, & \,0.03\, \\
\hline
\end{tabular}
\end{table}

Another impurity species relevant for the studied ICRF scenario is the second most abundant isotope of neon,
$^{22}{\rm Ne}^{10+}$ (natural abundance 9.3\%). In fact, for \mbox{$f=37~{\rm MHz}$} the IC resonance
of $^{22}{\rm Ne}$ impurities is located closer to the plasma center than the Be resonance
(see Table~\ref{tab1}).
As a~result, FW excited by a LFS antenna approaches the resonance of $^{22}{\rm Ne}$ ions first and hence
one can tune most of RF power to be absorbed by neon ions.
This is clearly illustrated in Fig.~4(b), where a~scan in $X_{^{22}{\rm Ne}}$ is made
for two beryllium background levels. E.g., for $X_{\rm Be}=1\%$, adding as
little as 0.02\% of $^{22}{\rm Ne}$ ions allows changing the dominant impurity
absorption channel and bringing the absorption region even closer to the plasma center.
Another stable isotope of neon, $^{21}{\rm Ne}$, is very rare (abundance 0.27\%), and for the considered conditions
not absorbing much of RF power.
A~preliminary conclusion one can draw from Figs.~\ref{fig4}(a) and (b) is that
having extra impurity species with a~similar \mbox{$Z/A \approx 0.43-0.45$}
and IC resonant layers located close to Be resonance ($\left| x_{\rm IC}^{\rm{(imp)}} - x_{\rm IC}^{\rm{(Be)}}\right|/a \simeq 0.1$,
see Table~\ref{tab1}) is not hindering the here considered heating scenario.

\noindent\textbf{\quad Comparison of the collisional power redistribution}

\begin{table}
\vspace{-2mm}
\caption{\label{tab2}
RF power split (direct heating and collisional redistribution) for the ($^3$He)-DT and (Be)-D-T ICRF scenarios.}
\begin{tabular}{|l|c|c|c|}
\hline
 & \scriptsize{Electrons} & \scriptsize{$^3$He/Be ions} & \scriptsize{Bulk ions} \\
\hline
a) ($^3$He)-DT scenario: & & & \\
\scriptsize{Direct RF heating}       & 9\% & 76\% & 14\% \\
\scriptsize{Collisional redistribution (10~MW)}    & 29\% & 5\%  & 62\% \\
\scriptsize{Collisional redistribution (20~MW)}    & 43\% & 4\%  & 50\% \\
\hline
b) (Be)-D-T scenario: & & & \\
\scriptsize{Direct RF heating}         & 6\% & 82\% & 12\% \\
\scriptsize{Collisional redistribution (10~MW)}    & 10\% & 9\%  & 81\% \\
\scriptsize{Collisional redistribution (20~MW)}    & 17\% & 7\%  & 76\% \\
\hline
\end{tabular}
\end{table}

Finally, we compare the collisional power redistribution for the $^3$He
and Be minority heating scenarios. As outlined in the introduction, the critical energy,
at which an equal amount of absorbed RF power collisionally goes to bulk ions and electrons,
increases with the atomic mass of the energetic particles:
$E_{\rm crit}[^{3}{\rm He}] \approx 25T_{e}$ and
$E_{\rm crit}[\Be] \approx 74T_{e}$, respectively.
Hence, at comparable ICRF power densities,
a stronger bulk ion heating is expected for the (Be)-D-T scenario.

In terms of the fraction of RF power absorbed directly by various plasma
species, the two ICRF scenarios are very similar.
As follows from Table~\ref{tab2}, minority ions ($^3$He and Be, respectively)
absorb about $75-80\%$ of the launched power,
$< 10\%$ of the RF power is deposited on electrons,
and the rest $10-15\%$ is absorbed by bulk D and T ions.
Using the RF power densities evaluated by TORIC,
the Fokker-Planck solver SSFPQL~\cite{ssfpql} was used
to estimate the collisional power redistribution between the species.
For $P_{\rm ICRF}=10~{\rm MW}$ of coupled RF power,
the fraction of bulk ion D-T heating is 62\%
for the ($^3$He)-DT scenario. This reduces to 50\% for 20~MW
of coupled RF power.
The electron heating fraction is computed to be 29\% and 43\%, respectively.
Note also that a finite amount of RF power eventually goes to heat minority $^3$He
and Be impurities.
The computed results are quite similar to those reported in Ref.~\cite{dumont2013}
(however, a higher plasma density and temperatures were used in those simulations).

For the (Be)-D-T scenario, the fraction of bulk ion heating is larger.
For 10~MW of coupled RF power, about 80\% eventually ends
up in bulk D-T ions and only 10\% in electrons. Increasing the coupled RF power
does not result in any significant reduction of $p_{i}$. For $P_{\rm ICRF}=20~{\rm MW}$,
fuel D and T ions still absorb 76\% of the coupled power and the fraction of electron heating
increases only up to 17\%.

Note that if extra natural neon (with natural abundances of isotopes)
is puffed into the plasma,
this can result in a~somewhat higher electron heating fraction.
For example, adding 0.5\% of natural neon gas
($\Delta Z_{\rm eff} \approx 0.5$)
brings only $\sim 0.05\%$ of $^{22}$Ne ions.
As~shown in Fig.~\ref{fig4}(b), $^{22}{\rm Ne}$ ions can absorb a~lot of RF power
even at such very low concentrations.
Subsequently, a tail of high-energy ions is expected to develop
in the $^{22}{\rm Ne}$ distribution function which
can collisionally transfer a~significant fraction of power to the electrons.
Such a more complicated multi-impurity option deserves
a~further detailed study (this scenario is also numerically more
challenging because of the multiple ion-ion hybrid layers present
in the plasma) and the corresponding results will be reported elsewhere.

\vspace{-5mm}
\section{Discussion and conclusions}
\vspace{-3mm}

A big advantage of the ($^3$He)-DT heating scenario is that it gradually converts to
$\omega=2\omega_{c \Tr}$ heating, without any change in the ICRF frequency. Thus, $^3$He is needed
only during the ramp-up phase of the pulse and its puff
can be switched off once the plasma temperature is high enough for second
harmonic tritium heating to become relatively strong. Good performance
of this scenario has been proven experimentally and if the availability
of $^3$He is not problematic, this scenario should be considered as the main option
for ITER.

The results of our studies show that there is a good backup option,
especially for the very early heating stage of the pulse.
The design of the ITER ICRF system currently envisages the use of two antennas~\cite{lamalle2009, wilson2015}.
With that, one could imagine several ICRF strategies for the ramp-up phase.
For example, an interesting option would be
to start a pulse with one antenna operating at 37--38~{\rm MHz} (or 40~MHz) and
selectively heat Be impurities. When the plasma pre-heating with the first
antenna gets efficient, one can switch on the second antenna at 53~MHz
and start $^3$He puffing to increase $T_{i}$ further.
The first antenna is then switched off and the ICRF system is configured
to launch the RF power at a higher frequency later on (e.g., either for
$\omega=2\omega_{c \Tr}$ heating or for current drive to extend the pulse length).
In any case, by using the first antenna at 37--40~MHz when $T_{i}$ is low,
one can reduce the total consumption of $^3$He during the ramp-up phase.

For ITER operating at full magnetic field \mbox{$B_0=5.3~{\rm T}$}, the proposed
(Be)-D-T ICRF scenario requires \mbox{$f \approx 37~{\rm MHz}$} to achieve heating close to the plasma center.
At first glance, this frequency is not ITER-relevant since the official ICRF range
is $f=40-55~{\rm MHz}$~\cite{lamalle2009, wilson2015}.
However, ITER RF generators are foreseen to deliver RF power
in the range \mbox{$f=35-65~{\rm MHz}$} \cite{kazarian2011}. Furthermore,
according to the computed ICRF antenna performance in ITER,
the reduction of the plasma coupling if operating at a lower frequency
is reasonably small. If compared to the results for the officially approved frequency 40~MHz,
the coupled power is lower by about $15\%$ only for $f=38~{\rm MHz}$ and by $20\%$ for 37~MHz
\cite{antiter, lamalle2009, itericrf}.

Yet, if the extension of the frequency range for ITER is not possible,
operating at $f=40~{\rm MHz}$ sets no major physics limitations
for the proposed (Be)-D-T scenario. A drawback of such operation is the non-central
power deposition, with the heating maximum shifted about $60~{\rm cm}$
to the high field side ($r/a \approx 0.3-0.35$).
Also for this configuration, one can expect a stronger effect of parasitic
alpha-particle absorption since at this frequency the IC resonance
of D and $^4$He ions is located centrally.
Direct absorption of ICRF power by energetic alpha particles is a potential showstopper for the (Be)-D-T heating
scenario at higher $T_{i}$.
Using the recently upgraded version of the TORIC code~\cite{toric2015}, we
have estimated that this absorption is reasonably small up to $T_{0} \simeq 10-15~{\rm keV}$:
at $T_{0}=10/15~{\rm keV}$ the fraction of alpha-particles in the plasma core is computed to be about $0.2\%/0.5\%$
and they absorb $p_{\alpha} \approx 7\%/19\%$ $(p_{\rm Be}=72\%/52\%)$ of the incoming RF power, respectively.
However, channeling of fusion alpha-particle power using minority ion catalysis \cite{fisch}
could potentially facilitate the application of Be impurity heating in D-T plasmas at higher $T_0$.

We want to outline a few examples of the three-ion ICRF heating
in D-T plasmas already reported in the literature: \\
(a) During the first D-T experiments in TFTR, reaching localized electron heating
and driving non-inductive current via mode conversion, was found to be unsuccessful.
The reason for that was attributed to the presence of lithium impurities in the plasma~\cite{wilson1998}.
Numerical computations showed that most of the RF power was absorbed by $^7{\rm Li}$ ions ($X_{^7 {\rm Li}} \approx 0.5\%$).
This absorption was considered as parasitic, and extensive wall conditioning
with $^6$Li-enriched pellets was used later to eliminate this effect and recover normal MC heating. \\
(b) For the JET pulse 42769, Start et al. reported $X_{\Be} \approx 1.5\%$.
It was estimated that the Be impurities were absorbing about 40\% of the RF power~\cite{start1999}.
Such a non-negligible Be heating was observed even though the IC resonance of Be impurities was located off-axis
and although the used D:T ratio $(X_{\De}=18\%)$ was quite different to the optimal values we have computed. \\
(c) In Ref.~\cite{dve2002} (Fig. 7), dominant absorption of RF power by Be ions in D:T=1:1 ITER plasmas
was computed for $X_{\Be}=1\%$, $f=40~{\rm MHz}$ and $k_{\|}^{\rm{(ant)}}=4~{\rm m}^{-1}$. \\
(d) In recent papers, where the ICRF scenarios for the activated phase of ITER were studied,
a significant impurity absorption, mostly by argon, was reported \cite{budny2012, dumont2013}.
Because the ICRF frequency was chosen to locate the $^3$He resonance centrally,
the argon resonance was placed at the very high field side edge. Note that fully
ionized argon ions have a charge-to-mass ratio ($Z/A = 0.45$) very similar to that of Be ions.
Again this impurity absorption was considered as a parasitic effect; however, in \cite{dumont2013} it was
outlined that this effect deserves further investigation with respect to the ICRF operation in ITER.

As follows from the examples shown above and using the reported results, one can conclude that
Be can be a very important species for future JET and ITER D-T experiments
involving ICRF heating. In view of ITER, this scheme can be tested during the
forthcoming DTE2 campaign at JET~\cite{kazakov.jet}. For JET operating at $B_{0}=3.6~{\rm T}$,
ICRF heating of Be impurities in D-T plasmas
requires the lowest frequency available for the A2 antennas \mbox{$f = 25~{\rm MHz}$}.

Finally, the proposed $N=1$ impurity heating scenario in D-T plasmas is relevant for DEMO
and future fusion reactors. In fact, there are several choices for the resonant absorber species in D-T plasmas.
If for any reason Be can not be used as a wall material for the tokamak-reactor DEMO,
this can be equally replaced with the other impurities with
a similar charge-to-mass ratio, viz. $^{7}{\rm Li}^{3+}$
(ionization energy for the ${\rm Li}^{2+} \rightarrow {\rm Li}^{3+}$ transition is $\mathcal{E}_{\rm ion.}=122~{\rm eV}$  \cite{wikipedia}),
$^{22}{\rm Ne}^{10+}$
($\mathcal{E}_{\rm ion.}=1.4~{\rm keV}$),
$^{40}{\rm Ar}^{17+}$ ($\mathcal{E}_{\rm ion.}=4.1~{\rm keV}$) and $^{40}{\rm Ar}^{18+}$ ($\mathcal{E}_{\rm ion.}=4.4~{\rm keV}$),
etc.
Note that the abundance of lithium is no problem for a fusion reactor,
and argon and neon are likely to be available for gas puffing as these noble gases can be used
for impurity seeding and controlling the deposition of the heat load in a machine.

Summarizing the arguments, the reported results show that
heavy intrinsic impurities with $1/3 < Z/A < 1/2$ can be an efficient absorber of the ICRF power
in D-T plasmas, and such a resonant impurity RF heating can be used for increasing the bulk ion temperature $T_{i}$
during the ramp-up phase of the plasma pulse.

\small{
\emph{\textbf{Acknowledgements.}
The authors are grateful to the anonymous referee for his/her valuable
suggestion to clarify the effect of extra impurity species.
This work has been carried out within
the framework of the EUROfusion Consortium and has received funding
from the Euratom research and training programme 2014-2018 under grant
agreement No 633053. The views and opinions expressed herein do not
necessarily reflect those of the European Commission.}
}


\vspace{-6mm}
\bibliographystyle{unsrt}

\end{document}